\documentclass[a4paper,onecolumn,accepted=2021-11-08]{quantumarticle}
\pdfoutput=1 
\makeatletter
\newcommand\extralabel[2]{{\edef\@currentlabel{\@currentlabel#2}\label{#1}}}
\makeatother
\usepackage{graphicx}
\usepackage{amsmath,amssymb,amsfonts}
\makeatletter
\renewcommand{\pod}[1]{\allowbreak\mathchoice
  {\if@display \mkern 18mu\else \mkern 8mu\fi (#1)}
  {\if@display \mkern 18mu\else \mkern 8mu\fi (#1)}
  {\mkern4mu(#1)}
  {\mkern4mu(#1)}
}
\usepackage{amsthm}
\usepackage{algpseudocode}
\usepackage{algorithm}
\usepackage{float}
\usepackage{thm-restate}
\usepackage{verbatim}
\usepackage{fancyvrb}
\usepackage{subfig}
\usepackage{url}
\usepackage[dvipsnames]{xcolor}
\usepackage{cancel}
\usepackage{hyperref}
\usepackage{listings}
\usepackage{comment}
\usepackage{printlen} 
\usepackage[numbers,sort&compress]{natbib} 

\makeatletter
\newcommand\thefontsize[1]{{#1 The current font size is: \f@size pt\par}}
\makeatother
\newcommand{\be}{\begin{equation}}
\newcommand{\ee}{\end{equation}}
\newcommand{\ba}{\begin{array}}
\newcommand{\ea}{\end{array}}
\newcommand{\bea}{\begin{eqnarray}}
\newcommand{\eea}{\end{eqnarray}}

\input{Qcircuit}

\newcommand{\calC}{{\cal C }}

\newcommand{\ZZ}{\mathbb{Z}}

\newcommand{\C}{{\cal C}}

\renewcommand{\P}{{\cal P}}

\newcommand{\Gate}[1]{\textsc{#1}}
\newcommand{\hgate}{\Gate{h}}
\newcommand{\zgate}{\Gate{z}}
\newcommand{\ygate}{\Gate{y}}
\newcommand{\xgate}{\Gate{x}}
\newcommand{\czgate}{\Gate{cz}}
\newcommand{\sgate}{\Gate{s}}
\newcommand{\tgate}{\Gate{t}}
\newcommand{\pgate}{\Gate{p}}
\newcommand{\idgate}{\Gate{i}}

\newcommand{\cnotgate}{\Gate{cnot}}

\newcommand{\swapgate}{\Gate{swap}}

\newtheorem{lemma}{Lemma}

\renewcommand{\sec}[1]{\hyperref[sec:#1]{Section~\ref*{sec:#1}}}
\newcommand{\ssec}[1]{\hyperref[ssec:#1]{Subsection~\ref*{ssec:#1}}}
\newcommand{\fig}[1]{\hyperref[fig:#1]{Fig.~\ref*{fig:#1}}}
\newcommand{\tab}[1]{\hyperref[tab:#1]{Table~\ref*{tab:#1}}}
\newcommand{\lem}[1]{\hyperref[lem:#1]{Lemma~\ref*{lem:#1}}}
\newcommand{\prop}[1]{\hyperref[prop:#1]{Proposition~\ref*{prop:#1}}}
\newcommand{\thm}[1]{\hyperref[thm:#1]{Theorem~\ref*{thm:#1}}}

\newcommand{\rev}[1]{{#1}} 

\begin{document}

\title{Clifford Circuit Optimization with Templates and Symbolic Pauli Gates}

\author{Sergey Bravyi$^1$, Ruslan Shaydulin$^2$, Shaohan Hu$^3$, and Dmitri Maslov$^1$}
\date{{\small $^1$IBM Quantum, IBM Thomas J. Watson Research Center, Yorktown Heights, NY 10598\\ 
$^2$Mathematics and Computer Science Division, Argonne National Laboratory, Lemont, IL 60439\\
$^3$JPMorgan Chase \& Co., New York, NY 10017}}

\maketitle

\begin{abstract}
The Clifford group is a finite subgroup of the unitary group generated by the Hadamard, the $\cnotgate$, and the Phase gates. This group plays a prominent role in quantum error correction, randomized benchmarking protocols, and the study of entanglement.  Here we consider the problem of finding a short quantum circuit implementing a given Clifford group element.  Our methods aim to minimize the entangling gate count assuming all-to-all qubit connectivity.  First, we consider circuit optimization based on template matching and design Clifford-specific templates that leverage the ability to factor out Pauli and $\swapgate$ gates.  Second, we introduce a symbolic peephole optimization method.  It works by projecting the full circuit onto a small subset of qubits and optimally recompiling the projected subcircuit via dynamic programming.  $\cnotgate$ gates coupling the chosen subset of qubits with the remaining qubits are expressed using symbolic Pauli gates.  Software implementation of these methods finds circuits that are only 0.2\% away from optimal for 6 qubits and reduces the two-qubit gate count in circuits with up to 64 qubits by 64.7\% on average, compared to the Aaronson--Gottesman canonical form \cite{Aaronson2004}.
\end{abstract}

\section{Introduction}
One of the central challenges in quantum computation is the problem of generating a short schedule of physically implementable quantum gates realizing a given unitary operation, otherwise known as the quantum circuit synthesis/optimization problem.  In this paper, we focus on a restricted class of quantum circuits belonging to the Clifford group, which is a subgroup of the group of all unitary transformations.  Clifford group elements play a crucial role in quantum error correction \cite{nielsen2002quantum}, quantum state distillation \cite{bravyi2005universal, knill2005quantum}, randomized benchmarking \cite{knill2008randomized, magesan2011scalable}, study of entanglement \cite{nielsen2002quantum, bennett1996mixed}, and, more recently, shadow tomography \cite{aaronson2020shadow, huang2020predicting}, to name some application areas.  Clifford group elements are important and frequently encountered subsets of physical-level and fault-tolerant quantum circuits; sometimes, an entire quantum algorithm can be a Clifford circuit (e.g., Bernstein--Vazirani \cite{nielsen2002quantum} \rev{and its generalizations~\cite{bravyi2018quantum}}).

\rev{
A special property of the Clifford group that plays the central role in many applications
is being a unitary $2$-design~\cite{dankert2009exact,cleve2015near}. 
It guarantees that a random uniformly distributed
element of the Clifford group has exactly the same second order moments as the Haar random
unitary operator. Thus random Clifford operators can serve as a substitute for Haar
random unitaries in any application that depends only on the second order moments.
However, in contrast to Haar random unitaries,
any Clifford operators admit an efficient implementation by a quantum circuit.
For example,  randomized benchmarking~\cite{knill2008randomized, magesan2011scalable} provides a scalable
fidelity metric for multi-qubit operations which is insensitive
to the state preparation and measurement errors. Randomized benchmarking works by 
measuring the decay rate of a signal generated by a sequence of random Clifford operators
of varying length. The $2$-design property ensures that the effective
noise model obtained after averaging over the Clifford group
is the depolarizing channel with a single unknown noise parameter.
As another example, classical shadows~\cite{huang2020predicting} 
provide a succinct classical
description of a multi-qubit quantum state that can be efficiently 
measured in an experiment without performing the full state tomography.
At the same time, a classical shadow determines many
physically relevant properties of a state such as expected values of
observables.
A classical shadow of a quantum state $\rho$
is obtained by repeatedly preparing a state $U\rho U^\dag$ with a
random Clifford operator $U$ and measuring each qubit  in the computational basis. 
The ability to realize a random element of the Clifford group
by a short quantum circuit plays the central role in the above examples. }

\rev{
Clifford circuits also serve  as a basis change transformation in quantum simulation algorithms. 
For example, simultaneous diagonalization of mutually commuting Pauli operators by
a Clifford basis change can reduce the circuit depth for simulating quantum chemistry Hamiltonians~\cite{van2020circuit}.
Another example is tapering off qubits for quantum simulations by identifying
Pauli-type symmetries of quantum chemistry Hamiltonians~\cite{bravyi2017tapering,setia2020reducing}. 
Such symmetry operators can be
mapped to single-qubit Pauli $Z$  by applying a suitable Clifford circuit
after which the respective qubits can be removed from the simulation.}

Earlier studies of the synthesis of $n$-qubit Clifford circuits resulted in the construction of asymptotically optimal (i.e., optimal up to a constant factor) implementations in the number of gates used.  Specifically, the canonical form introduced by Aaronson and Gottesman~\cite{Aaronson2004} accomplishes this using $\Theta\left(n^2/\log(n)\right)$ gates \cite{Ketan2008Optimal}.  
In contrast, in this paper we focus on the practical aspects of Clifford circuit optimization---our goal is to implement a given Clifford unitary by a circuit with the smallest possible number of entangling gates.  \rev{We focus on the minimization of the $\cnotgate$ gate count, drawing motivation from physical layer realizations where entangling gates come at a higher cost than the single-qubit gates, and ignore the connectivity constraints.  While in the worst-case scenario ignoring connectivity may lead to an $O(n)$ blowup in the $\cnotgate$ gate count or depth (consider the cost of implementation of the maximal-distance $\cnotgate(x_1;x_n)$ gate in a linear chain with $n$ qubits), known difference between the upper bound on the circuit depth between all-to-all and Linear Nearest Neighbor (LNN) architectures remains small.  Indeed, for all-to-all architecture the best known upper bound on the two-qubit gate depth is $\frac{10}{3}n+O(\log(n))$ (obtained by combining Lemma 8 in \cite{bravyi2021hadamard} with Corollary III.2.2 in \cite{goubaultdebrugiere2021reducing}, and noting that $\czgate$ gate layer can be implemented in depth $(n{-}1)$ or $n$ depending on whether $n$ is even or odd), and the best-known lower bound is $\Omega\left(\frac{n}{\log(n)}\right)$ (obtained by a slight modification of the counting argument employed in \cite{Ketan2008Optimal}).  In the LNN, upper bound is $9n$ \cite{bravyi2021hadamard}, and lower bound is $2n{+}1$ \cite{kutin2007computation}.  The above suggests that executing a (random) Clifford circuit in restricted architectures (LNN may often be embedded in other architectures) comes with a relatively small overhead.  We also note that our methods and algorithms can be straightforwardly modified to respect a restricted connectivity and target depth minimization rather than gate count minimization.}

Current approaches to the synthesis of exactly optimal Clifford circuits are prohibitively expensive even for small parameters: the largest number of qubits for which optimal Clifford circuits are known is six \cite{optimal6qubit}.  Using these exhaustive tools leaves little hope of scaling optimal implementations beyond six qubits.  Thus, efficient heuristics are desirable for practical applicability.  Here we focus on the synthesis and optimization of Clifford circuits that cannot be obtained optimally, namely, circuits with $n\,{>}\,6$ qubits.

Here we develop heuristic approaches for the synthesis and optimization of Clifford circuits.  Our algorithms and their implementation bridge the gap between nonscalable methods for the synthesis of exactly optimal Clifford circuits and the suboptimal (albeit asymptotically optimal) synthesis methods.  Our circuit synthesizer is based on the reduction of the tableau matrix representing Clifford unitary to the identity, while applying gates on both the input and output sides.  Our optimization approach is based on the extension and modification of two circuit optimization techniques: template matching \cite{Maslov2008tm3} and peephole optimization \cite{Prasad2006}.  To generate an optimized circuit for a specific Clifford unitary, we first compile it using the tableau representation and then apply the optimization techniques to the compiled circuit.  We note that the optimization techniques can be applied independently of the synthesizer considered in this paper.

The first optimization technique we develop is a Clifford-specific extension of the template matching method \cite{Maslov2008tm3}.  We discuss previous results on template matching in depth in \ssec{tm_prev}.  We introduce a three-stage approach that leverages the observation that in Clifford circuits Pauli gates can always be ``pushed'' to the end of the circuit without changing the non-Pauli Clifford gates (i.e., Hadamard, controlled-NOT, and Phase gates) and that all $\swapgate$ gates can be factored out of any quantum circuit by qubit relabeling.  We thus partition the circuit into ``compute,'' ``$\swapgate$,'' and ``Pauli'' stages by ``pushing'' Pauli and $\swapgate$ gates to the end of the circuit.  Next we optimize the ``compute'' stage using templates.  Then we optimize the ``$\swapgate$'' stage by exploiting the fact that a $\swapgate$ gate can be implemented at the effective cost of one entangling gate if it can be merged with a $\cnotgate$ or a $\czgate$ gate. 

The second technique we develop is symbolic peephole optimization. It is inspired by the peephole optimization method first introduced in the context of reversible computations \cite{Prasad2006}.  At each step, the symbolic peephole algorithm considers subcircuits spanning a small set of qubits (2 and 3 in this paper) by introducing symbolic Pauli gates (SPGs) to replace the two-qubit gates that entangle qubits in the chosen set with a qubit outside of it.  The resulting Clifford+SPG subcircuit is optimized via dynamic programming using a library of optimal circuits.

We numerically evaluate the proposed methods using two sets of benchmarks.  The first benchmark is based on the database of optimal Clifford circuits constructed in~\cite{optimal6qubit}.  We consider a selection of 1,003 randomly sampled 6-qubit Clifford unitaries, conditional on the optimal $\cnotgate$ gate implementation cost being higher than 4 (otherwise, it is easy to implement such a unitary optimally).  The set of tools developed in this work is able to recover an optimal (in terms of the $\cnotgate$ count) implementation for $97.9\%$ of the circuits, while producing circuits no more than one $\cnotgate$ away from the optimal count in the worst case.  Second, to evaluate the performance on ``large'' circuits, we consider a toy model of Hamiltonian evolution with a graph state Hamiltonian, defined as follows.  For a given graph with $n$ nodes, the Hamiltonian evolution performs the transformation $(\czgate\,{\cdot}\,\hgate)^t$, where $\czgate$ gates apply to graph edges, $\hgate$ gates apply to graph nodes (individual qubits), and $t$ is the evolution time.  At integer times, the evolution by such a Hamiltonian is described by a Clifford unitary. Implementing it as a circuit $\czgate\,{\cdot}\,\hgate$ repeated $t$ times turns out to be less efficient than implementing it by using the techniques reported here.  The methods we developed are evaluated on a collection of 2,264 circuits and shown to reduce the average $\cnotgate$ gate count by 64.7\% compared with the methods proposed by Aaronson and Gottesman in \cite{Aaronson2004}.  We make the full benchmark and the raw results available online~\cite{rawdata}.

The rest of the paper is organized as follows. We begin by briefly revisiting relevant concepts and defining the notations (\sec{background}). We next discuss previous results that our work is based on  (\ssec{tm_prev}, \ssec{ph_prev}).  Following this discussion, we describe the proposed methods (\sec{algorithms}), report numerical results, and evaluate the performance (\sec{results}).  We conclude with a short summary (\sec{conclusions}).

\section{Background}\label{sec:background}

We assume basic familiarity with quantum computing concepts, stabilizer formalism, and Clifford circuits. Below we briefly introduce relevant concepts and notations.  For detailed discussion, the reader is referred to \cite{nielsen2002quantum} and \cite{Aaronson2004}.

Clifford circuits (also known as stabilizer circuits) consist of Hadamard ($\hgate$), Phase ($\sgate$, also known as $\pgate$ gate), and controlled-NOT ($\cnotgate$) gates, as well as Pauli $\xgate$, $\ygate$, and $\zgate$ gates.  We use $I$ to denote the identity gate/matrix.
We also utilize the controlled-$\zgate$ ($\czgate$) gate, which can be constructed as a circuit with Hadamard and $\cnotgate$ gates as follows,

\begin{equation}
\raisebox{5mm}{
    \Qcircuit @C=0.5em @R=0.5em {
& \ctrl{1} & \qw & & & \qw & \ctrl{2} & \qw & \qw  \\
& & &  \push{\rule{.3em}{0em}\raisebox{-3mm}{=}\rule{.3em}{0em}} & & & & & \\
& \ctrl{-1} & \qw & & & \gate{\hgate} & \targ &\gate{\hgate} & \qw
}
}.
\label{eq:cz}
\end{equation}

Clifford circuits acting on $n$ qubits generate a finite group $\C_n$, known as the Clifford group.  An important property of Clifford circuits is that Clifford gates $\hgate$, $\sgate$, and $\cnotgate$ map tensor product of Pauli matrices into tensor products of Pauli matrices.  This property can be employed to ``push'' Pauli gates through the Clifford gates $\hgate$, $\sgate$, and $\cnotgate$ as follows:

\vspace{-5mm}
\begin{gather}
    \hgate\xgate = \zgate\hgate, \quad \hgate\ygate=-\ygate\hgate, \quad \hgate\zgate=\xgate\hgate, \quad \sgate\xgate = \ygate\sgate, \label{eq:paulipush0} \\
    \quad \sgate\ygate=-\xgate\sgate, \quad \sgate\zgate=\zgate\sgate, \label{eq:paulipush1} \\
    \cnotgate_{1,2}\xgate_1 = \xgate_1\xgate_2\cnotgate_{1,2}, \; \cnotgate_{1,2}\xgate_2 = \xgate_2\cnotgate_{1,2},\label{eq:paulipush2} \\ 
    \cnotgate_{1,2}\zgate_2 = \zgate_1\zgate_2\cnotgate_{1,2}, \; \cnotgate_{1,2}\zgate_{1} = \zgate_1\cnotgate_{1,2}.
    \label{eq:paulipush3}
\end{gather}

Our approach combines two building blocks:  Clifford-specific extension of template matching and symbolic peephole optimization.  Below we briefly review these techniques. While the developed methods reduce both single- and two-qubit gate count, in this paper we focus on the optimization of the number of two-qubit gates it takes to implement a Clifford group element.  The reason for our focus is that the leading quantum information processing technologies, trapped ions \cite{debnath2016demonstration} and superconducting circuits \cite{IBM}, both feature two-qubit gates that take longer time and have higher error rates compared with those of single-qubit gates.

\subsection{Template Matching}\label{ssec:tm_prev}

A size $m$ template \cite{Maslov2008tm3} is a sequence of $m$ gates that implements the identity function:

\vspace{-3mm}
\[
T = G_0G_1\ldots G_{m-1} = I.
\]

The templates can be used to optimize a target circuit as follows. First, a subcircuit \linebreak $G_iG_{i+1 \pmod m}\ldots G_{i+p-1 \pmod m}$ of the template is matched with a subcircuit in the given circuit.  If the gates in the target circuit can be moved together, this sequence of gates can be replaced with the inverse of the other $m{-}p$ gates in the template.  The larger the length $p$ of the matched sequence is, the more beneficial it is to perform the replacement, and for any $p\,{>}\,\frac{m}{2}$ the gate count is reduced.  The exact criteria for the application of the template depends on the choice of the objective optimization criteria (e.g., depth, total gate count, 2-qubit gate count).  More formally, for parameter $p$, $\frac{m}{2}{\leq} p{\leq} m$, the template $T$ can be applied in two directions as follows,

\begin{equation}
  \begin{aligned}[t]
    \text{\textbf{Forward: }} & G_iG_{i+1 \pmod m}\ldots G_{i+p-1 \pmod m} 
    \rightarrow G_{i-1 \pmod m}^\dagger G_{i-2 \pmod m}^\dagger \ldots G_{i+p \pmod m}^\dagger, \\
    \text{\textbf{Backward: }} & G_i^\dagger G_{i-1 \pmod m}^\dagger\ldots G_{i-p+1 \pmod m}^\dagger 
    \rightarrow G_{i+1 \pmod m}^\dagger G_{i+2 \pmod m}^\dagger \ldots G_{i-p \pmod m}^\dagger .
  \end{aligned}
  \label{eq:templateapp}
\end{equation}

Any template $T$ of size $m$ should be independent of smaller templates; that is,  an application of a smaller template should not decrease the number of gates in $T$ or make it equal to another template.  Circuit optimization using template matching is an iterative procedure where at each step we start at an index gate and attempt to match a given template by considering gates left to the index gate in the target circuit.  If the matched gates can be moved together and the substitution is beneficial, the template is applied as defined above.  This step is repeated by incrementing the position of the index gate by one when no match is found until the last gate is reached.

Circuit optimization with templates was originally proposed in~\cite{Maslov2008tm3}.  This work has been extended with the introduction of graph-based matching techniques~\cite{Raban2019}.  While the methods in these references are applicable to Clifford circuits since they are defined for universal quantum circuits, neither of them leverages the particular structure Clifford circuits have for optimization.
\rev{After completion of the present work we became aware
that template-based optimization techniques
have been recently applied to Clifford circuits in~\cite{sivarajah2020t}. 
}

\subsection{Peephole Optimization of Quantum Circuits}\label{ssec:ph_prev}

Peephole optimization \cite{Prasad2006} is an iterative local optimization technique that optimizes a circuit by considering subcircuits spanning small subsets of qubits $A$ and attempting to replace them with an optimized version drawn from a database (or synthesized on the spot in some other versions).  At each step, for a given gate all subcircuits on a fixed small number of qubits (e.g., $|A|{=}4$ in~\cite{Kliuchnikov2013ph}) including that gate are considered. For each subcircuit, its cost and the optimal cost (retrieved from the database of precomputed optimal circuits) of the unitary it implements are compared.  If a substitution is beneficial, the given subcircuit is replaced with its optimal implementation. The step is repeated for all gates until a convergence criterion is satisfied.  Peephole optimization of reversible circuits was introduced in~\cite{Prasad2006} and identified to be complementary to template matching.  Since its introduction in the context of reversible computations, this approach has been applied to Clifford circuits \cite{Kliuchnikov2013ph}.

The performance of the standard peephole optimization is limited by the need to store the entire database of optimal circuits in memory and to perform $O\left({n-2 \choose |A|-2} g^3\right)$ lookups, where $g$ is the number of gates in the circuit~\cite{Prasad2006}.  Furthermore, since the size of the $n$-qubit Clifford group (inclusive of the Pauli group) equals ${2^{2n+n^2}\prod\limits_{j=1}^{n}(2^{2j}{-}1)}$ and grows very quickly with $n$, it is unlikely that all optimal circuits can be found and stored in a suitable database for more than $6$ qubits \cite{optimal6qubit}. 

\section{Algorithms}\label{sec:algorithms}

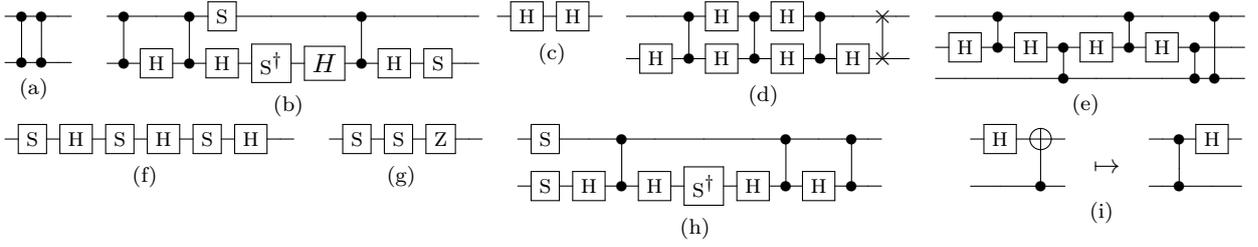
\begin{figure*}
\subfloat[\label{fig:tm1}]{
\Qcircuit @C=0.5em @R=1.5em {
         & \ctrl{1} & \ctrl{1} & \qw & \qw\\
         & \control\qw & \control\qw & \qw & \qw\\
     }
}
\quad
\subfloat[\label{fig:tm2}]{
    \Qcircuit @C=0.5em @R=0.5em {
         & \ctrl{1} & \qw & \ctrl{1} & \gate{\sgate} & \qw & \qw & \ctrl{1} & \qw & \qw & \qw & \qw\\                                                                                                            
         & \control\qw & \gate{\hgate} & \control\qw & \gate{\hgate} & \gate{\sgate^\dag} & \gate{H} & \control\qw & \gate{\hgate} & \gate{\sgate} & \qw & \qw\\
     }                      
}
\hspace{0.05in}\subfloat[\label{fig:tm3}]{
    \Qcircuit @C=0.5em @R=0.5em {
          & \gate{\hgate} & \gate{\hgate} & \qw &\\
     }                      
} 
\hspace{0.01in}\subfloat[\label{fig:tm4}]{
    \Qcircuit @C=0.5em @R=0.5em {
         & \qw & \ctrl{1} & \gate{\hgate} & \ctrl{1} & \gate{\hgate} & \ctrl{1} & \qw & \qswap & \qw & \qw\\                                                                                                          
         & \gate{\hgate} & \control\qw & \gate{\hgate} & \control\qw & \gate{\hgate} & \control\qw & \gate{\hgate} & \qswap \qwx[-1] & \qw & \qw\\
     }                      
}
\hspace{0.09in}\subfloat[\label{fig:tm5}]{
    \Qcircuit @C=0.5em @R=0.5em {
         & \qw & \ctrl{1} & \qw & \qw & \qw & \ctrl{1} & \qw & \qw & \ctrl{2} & \qw & \qw\\                                                                                                                 
         & \gate{\hgate} & \control\qw & \gate{\hgate} & \ctrl{1} & \gate{\hgate} & \control\qw & \gate{\hgate} & \ctrl{1} & \qw & \qw & \qw\\
         & \qw & \qw & \qw & \control\qw & \qw & \qw & \qw & \control\qw & \control\qw & \qw & \qw\\
     }                      
}
\vspace{-0.12in}
\\
\subfloat[\label{fig:tm6}]{
    \Qcircuit @C=0.5em @R=0.5em {
         & \gate{\sgate} & \gate{\hgate} & \gate{\sgate} & \gate{\hgate} & \gate{\sgate} & \gate{\hgate} & \qw & \qw\\
         }                      
}
\quad
\subfloat[\label{fig:tm7}]{
    \Qcircuit @C=0.5em @R=0.5em {
          & \gate{\sgate} & \gate{\sgate} & \gate{\zgate} & \qw & \qw\\
     }                      
}
\quad
\subfloat[\label{fig:tm8}]{
    \Qcircuit @C=0.5em @R=0.5em {
         & \gate{\sgate} & \qw & \ctrl{1} & \qw & \qw & \qw & \ctrl{1} & \qw & \ctrl{1} & \qw & \qw\\                                                                                                            
         & \gate{\sgate} & \gate{\hgate} & \control\qw & \gate{\hgate} & \gate{\sgate^\dag} & \gate{\hgate} & \control\qw & \gate{\hgate} & \control\qw & \qw & \qw\\
         }                      
}
\qquad\quad
\subfloat[\label{fig:tmhp}]{
\Qcircuit @C=0.5em @R=0.37em {
         & \gate{\hgate} & \targ & \qw & & & \qw & \control\qw & \gate{\hgate} &\qw \\                                    
         &    & \qwx      & & \push{\rule{.3em}{0em}\mapsto\rule{.3em}{0em}} & & & \qwx & \\
         & \qw & \ctrl{-1} &\qw & & & \qw & \ctrl{-1} & \qw & \qw\\
     }
}
\caption{ %
Templates \protect\subref{fig:tm1}--\protect\subref{fig:tm8} are used for template matching.  The rewriting rule \protect\subref{fig:tmhp} is used for Hadamard gate pushing.
}
\label{fig:templates}
\end{figure*}

We introduce two algorithms for Clifford circuit optimization and apply them to the problem of compiling optimized Clifford circuits.  The first algorithm is a Clifford-specific extension of the template matching technique, which we describe in \ssec{tm}.  The second algorithm is symbolic peephole optimization, detailed in \ssec{ph}.

These optimizations can be applied in at least the following two ways. First, if the input is a Clifford unitary, we begin by synthesizing a circuit using a ``greedy'' compiler (described in \ssec{baseline}) and then reduce the gate count by our proposed circuit optimization techniques.  Second, if the input is already a Clifford circuit, we can either resynthesize it or apply the circuit optimizations directly.  The gate count in the final circuit can be further decreased at the cost of increasing the runtime by a constant factor if the circuit is resynthesized $k$ times using a randomized version of the ``greedy'' compiler, the $k$ circuits are optimized individually, and the best of the $k$ results is picked. Note that the $k$ repetitions can be done in parallel.

\subsection{``Greedy'' Compiler}\label{ssec:baseline}

Suppose $U\,{\in}\,\C_n$ is a Clifford unitary to be compiled and $L\,{\in}\,\C_n$ is an operator that reproduces the action of $U$ on a single pair of Pauli operators, $\xgate_j$ and $\zgate_j$.  In other words, $UP U^{-1} = L P L^{-1}$ for $P\,{\in}\, \{\xgate_j,\zgate_j\}$.  The requisite operator $L$, as well as a Clifford circuit with $O(n)$ $\cnotgate$s implementing $L$, can be easily constructed for any given qubit $j$ by using the standard stabilizer formalism~\cite{Aaronson2004}.  Then the operator $L^{-1}U$ acts trivially on the $j$th qubit and can be considered as an element of the Clifford group $\calC_{n-1}$.  The greedy compiler applies this operation recursively such that each step reduces the number of qubits by one.  A qubit $j$, picked at each recursion step, is chosen such that the operator $L$ has the minimum $\cnotgate$ count.  In the randomized version of the algorithm, qubit $j$ is picked randomly.  The compiler runs in time $O(n^3)$ and outputs a circuit with the $\cnotgate$ count at most $3n^2/4+O(n)$.  We also developed and employ a bidirectional version of the greedy compiler that follows the same strategy as above except that each recursion step applies a transformation $U\gets L^{-1} U R^{-1}$, where $L, R \,{\in}\, \C_n$ are chosen such that after the transformation $U$ acts trivially on the $j$th qubit and the combined $\cnotgate$ count of $L$ and $R$ is minimized.  In \sec{results}, we use the bidirectional version of the greedy compiler as it leads to lower $\cnotgate$ costs of optimized circuits.  We include a detailed description of the greedy compilers in Appendix~A.

\subsection{Template Matching for Clifford Circuits}\label{ssec:tm}

We extend template matching, described in \ssec{tm_prev}, by introducing a three-stage approach that takes advantage of the observation that Clifford gates map tensor products of Pauli matrices into tensor products of Pauli matrices.  Below we describe the features used in the proposed three-stage approach.  In \ssec{fullalg}, we combine this approach with symbolic peephole optimization.

First, we partition the circuit into three stages, ``compute,'' ``$\swapgate$,'' and ``Pauli'', by pushing $\swapgate$ and Pauli gates to the end of the circuit.  Paulis are ``pushed'' according to the rules in Eqs.~(\ref{eq:paulipush0}, \ref{eq:paulipush1}, \ref{eq:paulipush2}, \ref{eq:paulipush3}).  This step results in the construction of the ``compute'' stage consisting of $\hgate$, $\sgate$, $\cnotgate$, and $\czgate$ gates only.

Second, we apply the template matching to the ``compute'' stage.  We further simplify template matching by converting all two-qubit gates into $\czgate$ gates (at the cost of introducing two Hadamard gates when the $\cnotgate$ is considered) before performing template optimization.  Templates are applied as described in \ssec{tm_prev}. The list of templates is given in Fig.~\ref{fig:tm1}-\ref{fig:tm8}.

We reduce the single-qubit gate count and increase the opportunities for template application by introducing Hadamard and Phase gate pushing.  Specifically, assuming that a circuit was optimized with templates, the idea is then to ``push'' Hadamard and Phase gates to one side of the two-qubit gates as far as possible. ``Pushing'' a gate through a two-qubit gate is implemented as the application of a template where a fixed subsequence must be matched. For example, the rule in Fig.~\ref{fig:tmhp} can be used to push a Hadamard to the right of the $\cnotgate$ gate. 

Note that once the circuit is optimized in terms of the two-qubit gate count, template matching can be applied to reduce the single-qubit gate count by restricting the set of templates and how they are applied.  This can be accomplished by applying templates spanning a single qubit and considering certain applications of templates with an even number of two-qubit gates.

Third, we consider $\swapgate$ gate optimization as a separate problem.  $\swapgate$ optimization is performed by observing that a $\swapgate$ gate can be implemented at the effective cost of one two-qubit gate if it is aligned with a two-qubit gate ($\cnotgate$ or $\czgate$) as, for example, in the following.
\vspace{-3mm}
\[
\Qcircuit @C=0.5em @R=0.5em {
& \ctrl{2} & \qswap     & \qw &                                       & & \ctrl{2} & \targ & \qw  \\
&          & \qwx       &  & \push{\rule{.3em}{0em}=\rule{.3em}{0em}} & & & \\
& \targ    & \qswap\qwx & \qw &                                       & & \targ & \ctrl{-2} & \qw
}
\]

In order to reduce the number of $\swapgate$s, the $\swapgate$ stage is resynthesized with the goal of aligning as many $\swapgate$s as possible with the two-qubit gates in the ``compute'' stage.

\subsection{Symbolic Peephole Optimization}\label{ssec:ph}

As outlined in \ssec{ph_prev}, various methods were proposed to create a database of optimal few-qubit Clifford circuits; some employ such databases to perform peephole optimization of larger Clifford circuits.  However, these methods are limited to few-qubit subcircuits that must be completely decoupled from the remaining qubits.  To address this limitation, we introduce a modified approach to Clifford circuit optimization, \emph{symbolic} peephole optimization. 

\begin{figure*}[t]
    \centering
    \includegraphics[width=\textwidth]{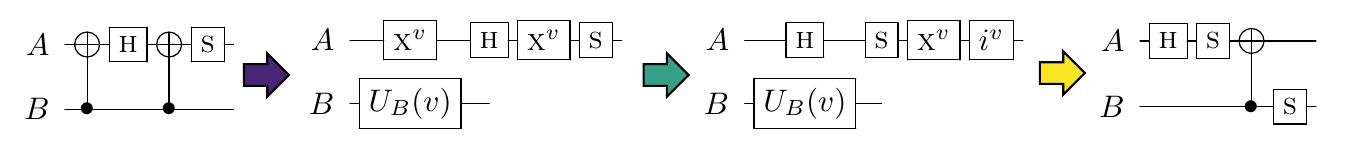}
    \caption{Example of symbolic peephole optimization.  {\em Purple arrow:} each entangling $\cnotgate$ gate is replaced with a symbolic Pauli gate (SPG) $\xgate^{v}$, where $v\in \{0,1\}$. Now the subcircuit acting on $A$ is isolated from the remaining qubits. 
    {\em Green arrow:} The subcircuit acting on $A$ is optimized to reduce the number of SPGs. Here we used the commutation rules $\hgate\xgate^v \,{=}\,\zgate^v\hgate$, $\xgate^v \zgate^v \,{=}\, (-i\ygate)^v$, and $\sgate\ygate^v\,{=}\,-\xgate^v\sgate$.
    {\em Yellow arrow:} The subcircuits acting on $A$ and $B$ are merged by replacing the SPG $\xgate^v$ with the $\cnotgate$.  The phase factor $i^v$ is replaced with the phase gate $\sgate$ acting on $B$.
    }
    \label{fig:PHexample}
\end{figure*}

Consider a circuit $U\,{\in}\,\mathcal{C}_n$ and a small subset of qubits $A\,{\subseteq}\,[n]$.  Our goal is to meaningfully define and optimize the restriction of $U$ onto $A$.  Let $B\,{=}\,[n]{\setminus}A$ be the complement of $A$. We say that a $\cnotgate$ gate is  \emph{entangling} if it couples $A$ and $B$. Assume without loss of generality that each entangling $\cnotgate$ has its target qubit in the set $A$ (otherwise, switch the control and the target by adding extra Hadamards). Partition entangling $\cnotgate$s into groups such that all $\cnotgate$s in the same group have the same control bit. Let $k$ be the number of groups. Expanding each entangling $\cnotgate$ as $\ket{0}\bra{0}\otimes I + \ket{1}\bra{1}\otimes X$, one can write
\[
U = \sum_{v\in\{0,1\}^k} U_A(v)\otimes U_B(v),
\]
where $U_A(v)$ is a Clifford circuit obtained from $U$ by retaining all gates acting on $A$ and replacing each entangling $\cnotgate$ from the $i$th group with the Pauli gate $X^{v_i}$ acting on the target qubit of the respective $\cnotgate$. Likewise, $U_B(v)$ is a (nonunitary) circuit obtained from $U$ by retaining all gates acting on $B$ and replacing each entangling $\cnotgate$ from the $i$th group with the projector $\ket{v_i}\bra{v_i}$ acting on the control qubit of the respective $\cnotgate$.  We refer to the single-qubit gates $\xgate^{v_i}$, $\ygate^{v_i}$, and $\zgate^{v_i}$ as \emph{Symbolic Pauli Gates (SPGs)}.  These are similar to controlled Pauli gates except that the control qubit is replaced by a symbolic variable $v_i \,{\in}\, \{0, 1\}$.

A symbolic Clifford circuit $U_A(v)$ can be optimized as a regular Clifford circuit on $|A|$ qubits with the following caveats. First, $U_A(v)$ must be expressed by using the Clifford+SPG gate set. The cost of $U_A(v)$ should be defined as the number of CNOTs plus the number of SPGs. Second, the optimization must respect the temporal order of SPGs. In other words, if $i{<}j$, then all SPGs controlled by $v_i$ must be applied before SPGs controlled by $v_j$. Third, the optimization must preserve the overall phase of  $U_A(v)$ modulo phase factors $(-1)^{v_j}$ or $i^{v_j}$. The phase factors can be generated by single-qubit gates $\zgate$ or $\sgate$ applied to control qubits of the entangling $\cnotgate$s. These conditions guarantee that the optimized circuit $U_A(v)$ can be lifted to a full circuit $U'\,{\in}\, \mathcal{C}_n$ that is functionally equivalent to $U$.  A toy optimization example is shown in Fig.~\ref{fig:PHexample}.

We  now describe the optimization of $U_A(v)$ in more detail. Let $\P_A$ and $\C_A$ be the groups of Pauli and Clifford operators acting on $A$, respectively.  The circuit $U_A(v)$ can be compactly specified by a $k$-tuple of Pauli operators $P_1,P_2,\ldots, P_k \in \P_A$ and a Clifford operator $R\,{\in}\, \C_A$ such that $U_A(v) = P_k^{v_k}\cdots P_2^{v_2}P_1^{v_1}R$ for all $v\in \{0,1\}^k$. Indeed, any SPG can be commuted to the left since Clifford gates map Pauli operators to Pauli operators.  The most general Clifford+SPG circuit that implements $U_A(v)$ can be parameterized as 
\begin{equation}
\label{eq:cspgform}
 U_A(v)=   (U_k^{-1} Q_k^{v_k} U_k)\cdots (U_2^{-1} Q_2^{v_2}U_2) (U_1^{-1} Q_1^{v_1}U_1) R
\end{equation}
for some Clifford operators $U_j\,{\in}\, \C_A$ and Pauli operators $Q_j=U_jP_j U_j^{-1}\in \P_A$.  The cost of the circuit in Eq.~\ref{eq:cspgform} includes the $\cnotgate$ count of subcircuits $U_j U_{j-1}^{-1}$ and the SPG count of controlled Pauli operators $Q_j^{v_j}$.  Note that $Q_j^{v_j}$ is a product of $|Q_j|$ single-qubit SPGs, where $|Q_j|$ is the Hamming weight of $Q_j$.  Denoting $U_0{:=} R^{-1}$, one can express the cost of the circuit in Eq.~(\ref{eq:cspgform}) as
\be
\label{f}
f=  \$(U_k)+\sum_{i=1}^k \$(U_i U_{i-1}^{-1}) 
   + \sum_{i=1}^k |U_i P_i U_i^{-1}|.
\ee
Here $\$(V)$ is the $\cnotgate$ cost of a Clifford operator $V\,{\in}\,\C_A$.  Our goal is to minimize the cost function $f$ over all $k$-tuples $U_1,U_2,\ldots,U_k\in \C_A$.  We claim that the global minimum of $f$ can be computed in time $O(k)$, as long as $|A|\,{=}\,O(1)$.  The key observation is that the function $f$ is a sum of terms that depend on at most two consecutive variables $U_i$ and $U_{i-1}$. Such functions can be minimized efficiently using the dynamic programming method; see, for example, \cite{aharonov2010efficient}.  Indeed, define intermediate cost functions $f_1,f_2,\ldots,f_k:  \C_A \to \ZZ_+$ such that $f_j$ is obtained from $f$ by removing the term $\$(U_k)$, retaining the first $j$ terms in the sums over $i$, and taking the minimum over $U_1,U_2,\ldots,U_{j-1}$.  More formally,
\be
\label{f1}
f_1(U_1) = \$(U_1 R) + |U_1 P_1 U_1^{-1}|
\ee
and
\[
f_j(U_j) = \min_{U_1,U_2,\ldots,U_{j-1}\in \C_A} \;\;
\sum_{i=1}^j \$(U_i U_{i-1}^{-1})  + \sum_{i=1}^j |U_i P_i U_i^{-1}|
\]
for $j=2,3,\ldots,k$.  Using the induction in $j$, one can easily check that 
\be
\label{f2}
f_{j}(U_{j}) =   |U_{j} P_{j}  U_{j}^{-1}| + \min_{U_{j-1}\in \C_A} \;  
f_{j-1}(U_{j-1}) + 
\$(U_{j} U_{j-1}^{-1})
\ee
for $j=2,3,\ldots,k$.  Below we assume that a lookup table specifying the $\cnotgate$ cost $\$(V)$ for all $V\,{\in}\, \C_A$ is available.  Then one can compute a lookup table of $f_1$ by iterating over all $U_1\,{\in}\, \C_A$ and evaluating the right-hand side of Eq.~(\ref{f1}). Proceeding inductively, one can compute a lookup table of $f_j$ with $j=2,3,\ldots,k$ by iterating over all $U_j\,{\in}\, \C_A$ and evaluating the right-hand side of Eq.~(\ref{f2}). Each step takes time roughly  $|\C_A|^2\,{=}\,O(1)$ since we assumed that $|A|\,{=}\,O(1)$.  Finally, use the identity
\be
\label{f3}
\min_{U_1,U_2,\ldots,U_k \in \C_A} f(U_1,U_2,\ldots,U_k) = \min_{U_k\in \C_A} \$(U_k) + f_{k}(U_k)
\ee
to compute the global minimum of $f$. Thus, the full computation takes time $O(k)$.

To make the above algorithm more practical, we exploited symmetries of the cost function, Eq.~(\ref{f}).  Namely, function $f$ is invariant under multiplying $U_j$ on the left by any element of the local subgroup  $\C_A^0\,{\subseteq}\,\C_A$ generated by the single-qubit gates $\hgate_a$ and $\sgate_a$ with $a\,{\in}\,A$.  In other words, $f(U_1,U_2,\ldots,U_k)$ depends only on the right cosets of the local subgroup $\C_A^{0}U_j$.  Thus one can restrict the minimizations in Eqs.~(\ref{f2},\ref{f3}) to some fixed set of coset representatives ${\cal R}\,{\subset}\,\C_A$ such that each coset $\C_A^{0}V$ has a unique representative $r(V){\,\in}\, {\cal R}$.  We chose $r(V)$ as the left-reduced form of $V$ defined in~\cite[Lemma 2]{optimal6qubit}.  This lemma provides an algorithm for computing $r(V)$ with the runtime $O(|A|^2)$.  Now each variable $U_i$ takes only $|{\cal R}|\,{=}\, |\C_A|/|\C_A^0|\,{=}\,|\C_A|/24^{|A|}$ values.  For example, $|{\cal R}|{=}20$ and $|{\cal R}|{=}6720$ for $|A|{=}2$ and $|A|{=}3$, respectively. Likewise, it suffices to compute the lookup table for the $\cnotgate$ cost $\$(V)$ only for $V\,{\in}\, {\cal R}$. This computation was performed using the breadth-first search on the Clifford group $\C_A$.

An important open question concerns the selection of the subsets $A$ to be considered.  From numerical experiments with $|A|\,{\in}\, \{2,3\}$, our most successful strategy turned out to be the random subset selection.  Specifically, we generate a list of all ${n \choose 2}$ pairs and ${n \choose 3}$ triples of qubits.  We run passes of the symbolic peephole method first on pairs of qubits and next on triples of qubits until no further improvement can be obtained.  At each pass of the symbolic peephole optimization, we randomly reshuffle both lists and run optimization on all the subsets in the reshuffled order.  We continue passes until either the optimal $\cnotgate$ count is reached (for circuits for which the optimal $\cnotgate$ count is known) or there is no improvement between two consecutive passes.

\subsection{Full Algorithm}\label{ssec:fullalg}

We combine the components described above in the following way. We begin by synthesizing the circuit using the ``greedy'' compiler described in \ssec{baseline}. Then the synthesized circuit is optimized as follows.  First, the circuit is partitioned into three stages.  Second, template matching and $\swapgate$ gate merging is performed until a pass yields no further optimization.  Third, symbolic peephole optimization is performed, as described in \ssec{ph}.  Lastly, a single pass of template matching is performed to reduce the single-qubit gate count.

\begin{figure}[H]
\centering
\begin{tabular}{p{0.3\textwidth} p{0.7\textwidth}}
{(a) The ratio of circuits for which the implemented methods recover optimal $\cnotgate$ count. The ``nonsmoothness'' of the line for the $\cnotgate$ count of 15 is due to only 3 circuits being considered.}
&
\raisebox{-5.5cm}{\includegraphics[width=0.66\textwidth]{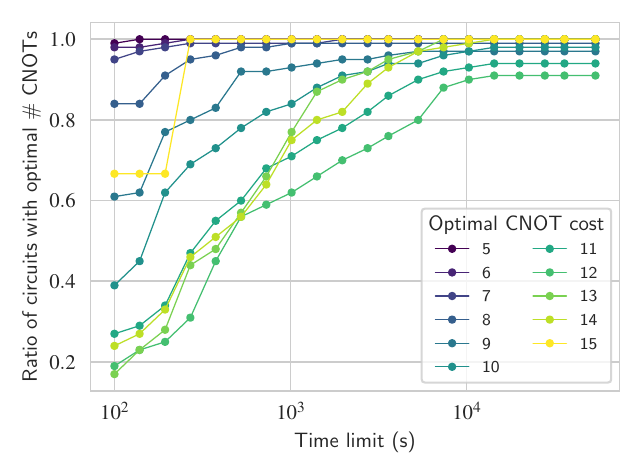}}
\\
(b) Mean running time of the implemented methods. Note that the actual running times are significantly lower than the time limit (black line) for most instances.
&
\raisebox{-5.5cm}{\includegraphics[width=0.66\textwidth]{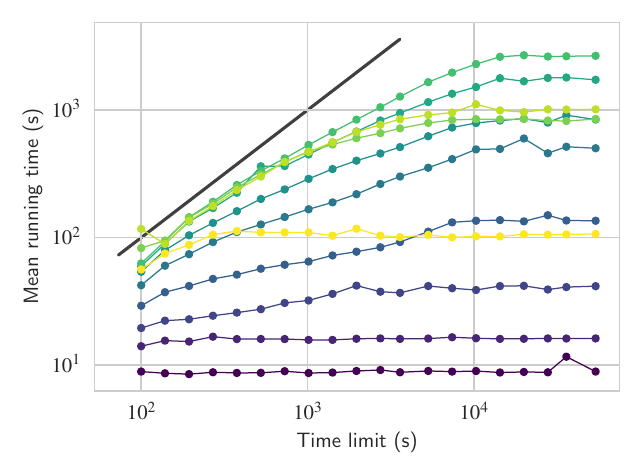}}
\end{tabular}
\caption{Quality of the solution (a) and the mean running time (b) for 6-qubit circuits with known optimal $\cnotgate$ gate count.  To demonstrate the trade-off between running time and the quality of the solution, we consider 20 time limits between 100 seconds and 15 hours. We observe that for all problems there exists a time limit at which the ratio of the recovered optimal circuits and the mean running time stops increasing.  This value depends on the hardness of the circuits; for the hardest circuits (optimal gate count of 12) the metrics are saturated at ${\approx}4$ hours. With this time limit we recover the optimal $\cnotgate$ count for 97.9\% of the circuits and observe the difference of 0.2\% between the average optimal $\cnotgate$ count and the average $\cnotgate$ count recovered by our software. Small deviations from monotonic growth of running time and quality with the time limit are due to the experiments being performed on a heterogeneous computing cluster and the random nature of the algorithm implementation. The mean running time being above the time limit for time limit $100$s is due to letting template matching complete even after the time limit is triggered.}
\label{fig:optimal_res}
\extralabel{fig:optimal_res:optimal_ratios}{a}
\extralabel{fig:optimal_res:optimal_runtimes}{b}
\end{figure}

\begin{table}[H]
\centering
\begin{tabular}{l|r|r|r|r|r|r|r|r|r}
Type & $n_q$ & $t_{\max}$ &  $C_{\text{orig}}$ &$C_{\text{A-G}}$ & $C_{\text{greedy}}$ & $C_{\text{opt}}$ & $r$ (\%) & \rev{$C_{\text{no }\swapgate}$} & \rev{$C_{\text{tket}}$} \\
\hline
 Path graph & 5 & 12 & 26.00 & 12.00 & 11.50 & 7.58 & 36.81 & \rev{4.50} & \rev{4.50} \\
 Path graph & 15 & 32 & 231.00 & 69.72 & 73.06 & 37.22 & 46.62 & \rev{22.50} & \rev{22.62} \\
 Path graph & 25 & 52 & 636.00 & 159.27 & 146.15 & 72.73 & 54.33 & \rev{43.50} & \rev{43.88} \\
 Path graph & 35 & 72 & 1241.00 & 270.79 & 224.67 & 109.78 & 59.46 & \rev{66.53} & \rev{67.17} \\
 Path graph & 45 & 92 & 2046.00 & 421.72 & 305.65 & 148.40 & 64.81 & \rev{89.96} & \rev{90.98} \\
 Path graph & 55 & 112 & 3051.00 & 589.71 & 384.27 & 188.29 & 68.07 & \rev{113.80} & \rev{115.46} \\
 Cycle graph & 5 & 10 & 27.50  & 19.60  & 12.90 & 7.80 & 60.20 & \rev{7.10} & \rev{7.30} \\
 Cycle graph & 15 & 30 & 232.50  & 163.17  & 76.00 & 45.90 & 71.87 & \rev{32.67} & \rev{33.60} \\
 Cycle graph & 25 & 50 & 637.50  & 424.60  & 155.56 & 88.12 & 79.25 & \rev{60.72} & \rev{61.94} \\
 Cycle graph & 35 & 70 & 1242.50  & 796.24  & 236.14 & 133.07 & 83.29 & \rev{89.10} & \rev{90.66} \\
 Cycle graph & 45 & 90 & 2047.50  & 1279.60  & 321.24 & 179.00 & 86.01 & \rev{118.01} & \rev{120.00} \\
 Cycle graph & 55 & 110 & 3052.50  & 1872.48  & 407.36 & 227.57 & 87.85 & \rev{147.04} & \rev{149.19} \\
 Square lattice & 4 & 4 & 10.00  & 6.00  & 5.00 & 3.50 & 41.67 & \rev{3.50} & \rev{3.50} \\
 Square lattice & 9 & 8 & 54.00  & 35.25  & 33.12 & 16.50 & 53.19 & \rev{15.38} & \rev{16.50} \\
 Square lattice & 16 & 12 & 156.00  & 95.42  & 93.17 & 40.08 & 57.99 & \rev{40.42} & \rev{45.33} \\
 Square lattice & 25 & 24 & 500.00  & 254.12  & 206.50 & 97.38 & 61.68 & \rev{89.75} & \rev{102.38} \\
 Square lattice & 36 & 36 & 1110.00  & 621.36  & 409.89 & 239.50 & 61.46 & \rev{221.39} & \rev{249.31} \\
 Square lattice & 49 & 16 & 714.00  & 858.44  & 553.12 & 320.19 & 62.70 & \rev{298.62} & \rev{343.44} \\
 Square lattice & 64 & 252 & 14168.00  & 2189.41  & 1237.07 & 887.38 & 59.47 & \rev{824.60} & \rev{904.80} \\
 Triangular lattice & 3 & 6 & 10.50  & 2.83  & 3.33 & 2.83 & 0.00 & \rev{2.83} & \rev{2.83} \\
 Triangular lattice & 6 & 10 & 49.50  & 22.60  & 18.60 & 9.70 & 57.08 & \rev{9.60} & \rev{10.20} \\
 Triangular lattice & 10 & 36 & 333.00  & 72.58  & 60.36 & 29.78 & 58.97 & \rev{26.42} & \rev{28.78} \\
 Triangular lattice & 15 & 90 & 1365.00  & 187.18  & 128.17 & 65.44 & 65.04 & \rev{61.40} & \rev{66.96} \\
 Triangular lattice & 21 & 24 & 562.50  & 303.08  & 211.67 & 117.33 & 61.29 & \rev{109.58} & \rev{117.79} \\
 Triangular lattice & 28 & 300 & 9481.50  & 726.63  & 418.46 & 272.20 & 62.54 & \rev{256.57} & \rev{272.00} \\
 Triangular lattice & 36 & 60 & 2562.00  & 928.92  & 556.60 & 356.87 & 61.58 & \rev{340.77} & \rev{369.03} \\
 Triangular lattice & 45 & 300 & 16254.00  & 1909.62  & 1023.72 & 801.87 & 58.01 & \rev{736.90} & \rev{767.44} \\
 Triangular lattice & 55 & 72 & 4927.50  & 2195.36  & 1229.04 & 968.68 & 55.88 & \rev{895.53} & \rev{922.10} \\
 Hexagonal lattice & 6 & 6 & 21.00  & 15.33  & 14.00 & 8.00 & 47.83 & \rev{7.50} & \rev{7.50} \\
 Hexagonal lattice & 24 & 24 & 375.00  & 285.88  & 197.25 & 101.83 & 64.38 & \rev{91.67} & \rev{101.58} \\
 Hexagonal lattice & 54 & 120 & 4356.00  & 1711.33  & 874.77 & 585.89 & 65.76 & \rev{539.67} & \rev{606.44} \\
 Heavy hexagon lattice & 12 & 12 & 78.00  & 79.50  & 58.92 & 28.42 & 64.26 & \rev{21.58} & \rev{22.67} \\
 Heavy hexagon lattice & 54 & 120 & 3630.00  & 1747.48  & 899.80 & 609.90 & 65.10 & \rev{561.50} & \rev{625.98} \\
\end{tabular}
\caption{Optimization results for Hamiltonian evolution circuits. For each graph on $n_q$ qubits, we generate and optimize $t_{\max}$ circuits corresponding to all integer numbers of steps between $1$ and $t_{\max} = \min(t_p,300)$. $C_{\text{orig}}$ is the average $\cnotgate$ gate count in the original circuits, $C_{\text{A-G}}$ is the average $\cnotgate$ gate count of the circuits in Aaronson--Gottesman canonical form \cite{Aaronson2004}, $C_{\text{greedy}}$ is the average $\cnotgate$ gate count of the circuits produced by the bidirectional ``greedy'' compiler, $C_{\text{opt}}$ is the average $\cnotgate$ gate count of the optimized circuits, and $r \,{=}\, (C_{\text{A-G}}{-}C_{\text{opt}})/C_{\text{A-G}}$ is the improvement in the average $\cnotgate$ gate count over the Aaronson--Gottesman canonical form. \rev{For all runs we set the time limit to 36 hours and stop both peephole optimization and template matching when the time limit is reached.}  We note that the ``greedy'' compiler by itself (without any further optimization) reduces the $\cnotgate$ gate count by $48.6\%$ compared to \cite{Aaronson2004}. \rev{We additionally compare the performance of our methods with the \texttt{CliffordSimp} method of tket framework~\cite{sivarajah2020t} applied to the output of our ``greedy'' compiler (column $C_{\text{tket}}$). As tket ignores the $\swapgate$ gates, we modified our implementation such that once the $\swapgate$s are factored out in template matching phase, they are ignored. The $\cnotgate$ gate counts are presented in column $C_{\text{no }\swapgate}$. We observe that our optimizations result in $\cnotgate$ counts that are $6.58\%$ lower on average as compared to tket, with larger improvements (up to $17.8\%$) observed for harder (deeper) circuits.}
}
\label{tab:hcz_bench}
\end{table}

\section{Experimental Results}\label{sec:results}

We ran two sets of computational experiments designed to test the performance of our synthesis and optimization algorithms, detailed in the next two subsections.  In addition, we compared our results to \cite{sivarajah2020t} as well as to 8-qubit $\tgate$ gate free circuits from \cite{Duncan2020}.  The comparison to \cite{sivarajah2020t} is detailed in \tab{hcz_bench} and the comparison to  \cite{Duncan2020} reads $24.4529$ (obtained using $10{,}000$ random samples) to $50{+}$ in \cite[Figure 3]{Duncan2020}.

\subsection{Recovering Optimal CNOT Count for Clifford Unitaries on Six Qubits}\label{sec:recover_optimal}

First we compare the proposed heuristic methods with the optimal Clifford compiler for $n\,{\le}\,6$ qubits \cite{optimal6qubit}.  The latter uses breadth-first search on the Clifford group to construct a database specifying the optimal $\cnotgate$ gate count of each Clifford operator.  As shown in \cite{optimal6qubit}, the optimal $\cnotgate$ gate count for $6$-qubit Clifford operators takes values $0,1,\ldots,15$.  We generate 1,003 uniformly sampled random Clifford unitaries with the $\cnotgate$ gate counts between 5 and 15.  We consider only unitaries with the $\cnotgate$ gate count ${\geq}5$ because one needs at least 5 $\cnotgate$s to entangle all 6 qubits.  For the $\cnotgate$ gate counts from 5 to 14, we consider 100 circuits for each cost value. For the $\cnotgate$ gate count of $15$, there are only 3 Clifford circuits (modulo single-qubit Cliffords on the left and on the right and modulo qubit permutations) to consider \cite{optimal6qubit}.

For each Clifford unitary, we start by synthesizing it using the bidirectional ``greedy'' compiler. The optimization is run as described in \ssec{fullalg}. The circuit is then resynthesized by using the randomized version of the compiler, and the resynthesized circuit is optimized. This process is repeated until the time limit is reached, and the circuit with the lowest $\cnotgate$ count is chosen as the output. Note that we also stop the peephole optimization when the time limit is reached, but we allow template matching to complete. The reason is that template matching is fast as compared to peephole optimization and allowing it to complete results in the actual running time above the time limit by only $0.66\%$ of the instances considered.

The quality of the solution obtained by the implemented methods as a function of the time limit is shown in \fig{optimal_res:optimal_ratios}. Our algorithm converges before exhausting the time limit on most instances. \fig{optimal_res:optimal_runtimes} shows actual observed mean running time as a function of the time limit.  We note that the combination of the iterative nature of symbolic peephole optimization and the randomized resynthesis allows the user to trade off the quality of the optimization and the running time as desired.

\subsection{Circuits for Hamiltonian Evolution}

To evaluate the performance of the proposed methods on circuits with $n{>}6$ qubits, we consider a toy model of Hamiltonian time evolution.  Suppose $G\,{=}\,(V,E)$ is a fixed graph with $n$ vertices.  We place a qubit at each vertex of $G$.  Define a Hamiltonian evolution circuit with time $t$ as 
\[
\left(  \prod_{(i,j)\in E} \czgate_{i,j}
\prod_{j=1}^n \hgate_j
\right)^t.
\]
The layers of Hadamard and $\czgate$ gates model time evolution under an external magnetic field and nearest neighbor two-qubit interactions, respectively.  We consider several choices for the interaction graph. First, we take instances of the path and cycle graphs with the number of qubits $n \in \{5,15,25,35,45,55\}$.  Second, we include all three regular plane tessellations (by triangles, squares, and hexagons).  We choose the numbers of vertices between $6$ and $64$ such that the convex hull spanned by the centers of masses of individual tiles in the gapless regular tiling is congruent to the basic tile.  Third, we consider a heavy hexagon grid, obtained from hexagonal tessellation by adding a node in the middle of each edge.  This set includes some of the frequently appearing qubit-to-qubit connectivities/architectures.  We consider the number of layers $1 \,{\leq}\, t \,{\leq}\, t_{\max}=\min(t_p,300)$, where $t_p$ is the period such that the Hamiltonian evolution with the number of layers $t_p$ produces the identity transformation. For each interaction graph $G$ we compute the $\cnotgate$ gate count of optimized circuits averaged over the number of layers $t\,{=}\,1,2,\ldots,t_{\max}$.  The total number of circuits considered is 2,264.  We set the time limit to 36 hours and stop both peephole optimization and template matching when the time limit is reached, only allowing the current pass of template matching to complete.  Allowing the current pass of template matching to complete results in only a small ratio of problems ($4.5\%$) to significantly (${\geq}10\%$) exceed the time limit.  The results are reported in \tab{hcz_bench}.  The maximum graph size in these experiments is $n{=}64$ because we represent $n$-qubit Pauli operators by a pair of $64$-bit integers in our C++ implementation; this limitation can be easily removed by revising the data structure.

\section{Conclusion}\label{sec:conclusions}

We reported a bidirectional synthesis approach and two circuit optimization techniques that extend known approaches to Clifford circuits by exploiting the unique properties of the Clifford group.  We demonstrate the effectiveness of these methods by recovering optimal $\cnotgate$ gate count for $98.9\%$ 6-qubit circuits (over 1,003 samples) and by reducing the $\cnotgate$ gate count by 64.7\% on average for Hamiltonian evolution circuits with up to 64 qubits (2,264 circuits considered), compared to Aaronson--Gottesman canonical form~\cite{Aaronson2004}.  We show evidence of the improvement in the gate count by a factor of 2 compared to other techniques, such as \cite{Duncan2020}.

\section*{Appendix A}

In this section, we describe greedy synthesizers that we developed to obtain initial circuits implementing the Clifford unitaries considered.  There are two versions of the synthesizer: unidirectional and bidirectional.  

Let us start with the unidirectional version.  Suppose  $C$ is an $n$-qubit Clifford operator.  The  greedy compiler sequentially constructs Clifford circuits $L_1,L_2,\ldots,L_n$ such that $C=L_1 L_2 \cdots L_j C_j$, where $C_j$ acts trivially on the first $j$ qubits. The circuit $L_j$ has the $\cnotgate$ cost $\$(L_j)\le 3(n{-}j)/2+O(1)$.  By assumption, $C_n$ acts trivially on all $n$ qubits, that is, $C_n$ is proportional to the identity.  In other words, $C= \omega L_1 L_2\cdots L_n$ for some irrelevant phase factor $\omega$. This gives a circuit implementing $C$ with the $\cnotgate$ cost 
\[
\sum_{j=1}^n \$(L_j) \le \sum_{j=1}^n \left( 3(n-j)/2 + O(1) \right) = 3n^2/4 + O(n).
\]
The desired circuits $L_j$ are constructed using the following lemma.
\begin{lemma}
\label{lemma:pair1}
There exists an algorithm that takes as input anti-commuting $n$-qubit Pauli operators $O$ and $O'$ and outputs a Clifford circuit $L\,{\in}\, \calC_n$ such that 
\be
\label{disentangler}
L^{-1} OL = \xgate_1 \quad \mbox{and} \quad L^{-1} O'L=\zgate_1.
\ee
The circuit $L$ has the $\cnotgate$ cost $\$(O,O'):=\$(L) \le 3n/2+O(1)$.  The algorithm has runtime $O(n)$.
\end{lemma}

We refer to a Clifford operator $L$ satisfying Eq.~(\ref{disentangler}) as a disentangler for the pair $(O, O')$.

To implement a Clifford operator $C$ as a circuit, apply the lemma to Pauli operators $O\,{=}\,C\xgate_1 C^{-1}$ and $O'\,{=}\,C\zgate_1 C^{-1}$. Let $L$ be the corresponding disentangler and $C_1\,{=}\,L^{-1} C$.  From Eq.~(\ref{disentangler}) we obtain $C_1 \xgate_1\,{=}\, \xgate_1 C_1$ and $C_1 \zgate_1 \,{=}\, \zgate_1 C_1$.  This means that $C_1$ commutes with all Pauli operators on the first qubit, which is possible only if $C_1$ acts trivially on the first qubit. We can now apply the lemma to $C_1$.  Ignoring the trivial action of $C_1$ on the first qubit, one can regard $C_1$ as an element of the smaller Clifford group $\calC_{n-1}$, thereby reducing the number of qubits that need to be considered by one.  After $n$ applications of the lemma we obtain the desired disentanglers $L_1,L_2,\ldots,L_n$.

\begin{proof}[\bf Proof of Lemma~\ref{lemma:pair1}]	
Let us say that Pauli operators $O$ and $O'$ are in the standard form if their action on any qubit $j$ falls into one of  the five cases shown below.
	\begin{center}
		\begin{tabular}{|c|c|c|c|c|c|}
			\hline 
			Case & $A$ & $B$ & $C$ & $D$ & $E$ \\
			\hline 
			\hline
			$O_j$ & $\xgate$ & $\xgate$ & $\xgate$ & $\idgate$ & $\idgate$ \\
			\hline
			$O_j'$ & $\zgate$  & $\xgate$ & $\idgate$ & $\zgate$ & $\idgate$ \\
			\hline
		\end{tabular}
	\end{center}
Recall that the single-qubit Clifford group $\calC_1$ acts by permutations on the Pauli operators $\xgate,\ygate,$ and $\zgate$. Thus one can transform any Pauli pair $(O,O')$ into the standard form by applying a layer of single-qubit Clifford operators.  This gives rise to a partition of $n$ qubits into five disjoint subsets $A, B, C, D,$ and $E$.  Note that $A$ has odd size since otherwise $O$ and $O'$ would commute.  Let $A(j)$ be the $j$-th qubit of $A$.  Consider the following circuit.

{\centering
	\begin{minipage}{1.0\linewidth}	
	\begin{algorithm}[H]
		\caption{Disentangling circuit}
		\begin{algorithmic}[1]
			\If{$1\notin A$}
			\State{$\swapgate_{1,A(1)}$}
			\EndIf \Comment{Now $O_1=\xgate$ and $O_1'=\zgate$}
			\For{$j\in C$} \Comment{Clean $O$ from $C$}
			\State{$\cnotgate_{1,j}$}
			\EndFor
			\For{$j\in D$} \Comment{Clean $O'$ from $D$}
			\State{$\cnotgate_{j,1}$ }
			\EndFor
			\State{$i\gets$ first qubit of $B$}
			\For{$j\in B\setminus \{i\}$} \Comment{Clean $(O,O')$ from $B{\setminus} \{i\}$}
			\State{$\cnotgate_{i,j}$ }
			\EndFor
			\State{$\cnotgate_{i,1}\hgate_i \cnotgate_{1,i}$ \Comment{Clean $(O,O')$ from $i$}}
			\State{$k\gets (|A|-1)/2$} \Comment{Due to anticommutativity, $|A|$ is odd; group all but qubit 1 in pairs}
			\For{$j=1$ to $k$ }\Comment{Clean $(O,O')$ from $A{\setminus}\{1\}$}
			\State{$\cnotgate_{A(2j+1),A(2j)}$}
			\State{$\cnotgate_{A(2j),A(1)}$}
			\State{$\cnotgate_{A(1),A(2j+1)}$}\Comment{Perform simultaneous mapping $\xgate\xgate\xgate\mapsto\xgate\idgate\idgate$ and $\zgate\zgate\zgate\mapsto\zgate\idgate\idgate$}
			\EndFor
		\end{algorithmic}
	\end{algorithm}
\end{minipage}
}

Let $L$ be the operator realized by the above circuit combined with the initial layer of single-qubit Cliffords.  Direct inspection shows that $L$ has the desired property, Eq.~(\ref{disentangler}), up to sign factors.  The latter can be fixed by applying Pauli $\xgate_1$ or $\ygate_1$ or $\zgate_1$ as the first gate of $L$.  Simple algebra shows that $L$ has the $\cnotgate$ gate count of at most $(3/2)|A| + |B|+|C|+|D|+O(1)\le 3n/2+O(1)$.
\end{proof}

Below we use the notation $\$(O,O')$ for the $\cnotgate$ gate count of the disentangling circuit constructed in Algorithm~1.

Our implementation of the greedy compiler optimizes the order in which the qubits are disentangled.  Namely, suppose that at some step $j$ of the compiler a subset of qubits $S_j$ has been disentangled such that $|S_j|\,{=}\,j$ and $C=L_1 L_2\cdots L_j C_j$, where $C_j$ acts trivially on $S_j$.  Let 
\[
p=\arg \min_{p\notin S_j} \$(C_j \xgate_p C_j^{-1}, C_j\zgate_p C_j^{-1})
\]
be a qubit with the smallest disentangling cost.  Let $L$ be the circuit disentangling $C_j \xgate_p C_j^{-1}$ and $C_j\zgate_p C_j^{-1}$. Set $S_{j+1}=S\cup \{p\}$ and $L_{j+1}=L\cdot \mathsf{SWAP}_{1,p}$.  Then $C=L_1L_2\cdots L_{j+1} C_{j+1}$, where $C_{j+1}$ acts trivially on $S_{j+1}$.  Thus one can proceed inductively.  The extra $\swapgate$ gates are isolated and incorporated back into the compiled circuit as described in \ssec{tm}.

The greedy synthesizer described above has the runtime $O(n^3)$.  Indeed, consider the first step of the synthesis.  Since the disentangling cost $\$(O,O')$ can be computed in time $O(n)$, picking a qubit with the smallest disentangling cost takes time $O(n^2)$. Computing the disentangling circuit $L$ and the product $L^{-1}C$ takes time $O(n^2)$ since $L$ contains $O(n)$ gates and the action of a single gate can be simulated in time $O(n)$ using the stabilizer formalism~\cite{Aaronson2004}.  Thus the full runtime of the greedy synthesizer is $O(n^3)$.

The bidirectional greedy synthesizer sequentially constructs Clifford circuits $L_1,L_2,\ldots,L_n$ and \linebreak $R_1,R_2,\ldots,R_n$ such that 
\[
C=(L_1 L_2\cdots L_j) C_j (R_j \cdots R_2 R_1),
\]
where $C_j$ acts trivially on the first $j$ qubits.  This gives a circuit implementing $C$ with the $\cnotgate$ cost at most
\[
\sum_{j=1}^n \$(L_j) + \$(R_j).
\]
Consider the first step, $j{=}1$.  Let us construct the circuits $L\,{=}\,L_1$ and $R\,{=}\,R_1$ (all subsequent steps are analogous).  Our goal is to minimize the combined cost $\$(L)+\$(R)$ subject to the constraint that $C_1=L^{-1} C R^{-1}$ acts trivially on the first qubit.  Equivalently, $C_1$ should commute with the Pauli operators $\xgate_1$ and $\zgate_1$.  Define $n$-qubit Pauli operators
\[
P:=R^{-1} \xgate_1 R, \;
P':=R^{-1} \zgate_1 R, \;
O := CP C^{-1}, \text{ and }
O' := CP' C^{-1}.
\] 
By definition, $R^{-1}$ is a disentangler for the pair $(P, P')$. Simple algebra shows that $C_1$ commutes with the Pauli operators $\xgate_1$ and $\zgate_1$ if and only if $L$ is a disentangler for the pair $(O,O')$.  The above shows that minimizing the combined cost $\$(L)+\$(R)$ subject to the constraint that $C_1=L^{-1} C R^{-1}$ acts trivially  on the first qubit is equivalent to minimizing the function
\[
f(P,P') = \$(P,P') + \$( CP C^{-1}, CP' C^{-1})
\]
over all pairs of $n$-qubit anti-commuting Pauli operators $P$ and $P'$.  Note that $f(P,P')$ can be computed in time $O(n)$ for a given pair $(P,P')$.  Once the optimal pair $(P,P')$ is found, one chooses $L$ and $R^{-1}$ as disentanglers for the Pauli pairs $(O,O')$ and $(P,P')$, respectively.  Since the total number of $n$-qubit anti-commuting Pauli pairs grows exponentially with $n$, the global minimum of $f(P,P')$ cannot be computed exactly for large $n$. To make the problem tractable, we restricted the minimization to Pauli operators $P,P'$ with weight at most two.  The number of such pairs $(P,P')$ is at most $O(n^3)$ since the anti-commutativity condition implies that the supports of $P$ and $P'$ must overlap on at least one qubit.  Now the minimum of $f(P,P')$ can be computed in time $O(n^4)$, and thus the full runtime of the compiler is $O(n^5)$.  Note that the unidirectional greedy compiler described earlier corresponds to $R{=}I$, that is, $P\,{=}\,\xgate_1$ and $P'\,{=}\,\zgate_1$.  Thus the bidirectional compiler subsumes the unidirectional one, even with the restricted minimization domain.

\section*{Disclaimer}

This paper was prepared for information purposes with contributions from the Future Lab for Applied Research and Engineering (FLARE) Group of JPMorgan Chase \& Co. and its affiliates, and is not a product of the Research Department of JPMorgan Chase \& Co. JPMorgan Chase \& Co. makes no explicit or implied representation and warranty, and accepts no liability, for the completeness, accuracy or reliability of information, or the legal, compliance, tax or accounting effects of matters contained herein. This document is not intended as investment research or investment advice, or a recommendation, offer or solicitation for the purchase or sale of any security, financial instrument, financial product or service, or to be used in any way for evaluating the merits of participating in any transaction.

\section*{Acknowledgments}

This work was supported in part by the U.S.\ Department of Energy (DOE), Office of Science, Office of Advanced Scientific Computing Research AIDE-QC and FAR-QC projects and by the Argonne LDRD program under contract number DE-AC02-06CH11357. Clemson University is acknowledged for generous allotment of compute time on the Palmetto cluster. \rev{We gratefully acknowledge the computing resources provided on Bebop, a high-performance computing cluster operated by the Laboratory Computing Resource Center at Argonne National Laboratory.} SB is partially supported by the IBM Research Frontiers Institute.

\bibliographystyle{plainnat}
\bibliography{coptbib}

\end{document}